\documentclass{jaa}
\usepackage{natbib}
\usepackage[colorlinks=true,citecolor=blue]{hyperref}
\usepackage{amsmath}
\usepackage{graphicx}

\def\deg{\ifmmode^\circ\else$^\circ$\fi}

\newcommand{\mic}{\,$\mu$m}

\pdfoutput=1

\begin{document}\sloppy

%%paper title
%%For line breaks \\ can be used within title
\title{Star-forming site RAFGL 5085: Is a perfect candidate of hub-filament system ?}
\author{L.~K. Dewangan\textsuperscript{1}, N.~K.~Bhadari\textsuperscript{1,2}, A.~K. Maity\textsuperscript{1,2}, Rakesh Pandey\textsuperscript{1}, Saurabh Sharma\textsuperscript{3}, T. Baug\textsuperscript{4}, and C. Eswaraiah\textsuperscript{5}}
%, Sachindra Naik\textsuperscript{1}, and Neeraj Kumari\textsuperscript{1,2}}
\affilOne{\textsuperscript{1}Astronomy and Astrophysics Division, Physical Research Laboratory, Navrangpura, Ahmedabad - 380009, India.\\}
\affilTwo{\textsuperscript{2}Indian Institute of Technology, Gandhinagar, Gujarat, India.\\}
\affilThree{\textsuperscript{3}Aryabhatta Research Institute of Observational Sciences (ARIES),
Manora Peak, Nainital 263 002, India.\\}
\affilFour{\textsuperscript{4}Satyendra Nath Bose National Centre for Basic Sciences, Block-JD, Sector-III, Salt Lake, Kolkata-700 106, 
India.\\}
\affilFive{\textsuperscript{5}Indian Institute of Science Education and Research (IISER) Tirupati, Rami Reddy Nagar, Karakambadi Road, Mangalam (P.O.), Tirupati 517 507, India.}

%%author names are separated by comma (,)
%%use \and before the last author name
%%use a * along with the number separated by comma
%% for the  author for correspondence
%%\textsuperscript{number} is used for affiliation
%%\affilOne, \affilTwo etc., upto \affilTwentyfive is possible
%%Please note the first letter after \affil is capitalised in the command
%%
%
%\author{AUTHOR1\textsuperscript{1}, AUTHOR2\textsuperscript{1} and AUTHOR3\textsuperscript{2,*}}
%\affilOne{\textsuperscript{1}Department of P, University X, Place Pincode, Country.\\}
%\affilTwo{\textsuperscript{2}Department of Q, University Z, Place Pincode, Country.}
%
%%escape two column mode for title, affiliation and abstract
%%by giving \twocolumn command as shown

\twocolumn[{

\maketitle

%%include \corres to print the corresponding author Email id
\corres{loku007@gmail.com, lokeshd@prl.res.in}

%%include \msinfo for
%%manuscript information such as
%%received, revised and accepted dates
%%
\msinfo{***}{***}

%%abstract
\begin{abstract}
To investigate the star formation process, we present a multi-wavelength study of a massive star-forming site RAFGL 5085, which has been associated with the molecular outflow, H\,{\sc ii} region, and near-infrared cluster. The continuum images at 12, 250, 350, and 500 $\mu$m show a central region (having M$_{\rm clump}$ $\sim$225 M$_{\odot}$) surrounded by five parsec-scale filaments, revealing a hub-filament system (HFS). In the {\it Herschel} column density ($N({{\rm{H}}}_{2})$) map, filaments are identified with higher aspect ratios (length/diameter)
and lower $N({{\rm{H}}}_{2})$ values ($\sim$0.1--2.4 $\times$10$^{21}$ cm$^{-2}$), while the central hub is found with a lower aspect ratio and higher $N({{\rm{H}}}_{2})$ values ($\sim$3.5--7.0 $\times$10$^{21}$ cm$^{-2}$). The central hub displays a temperature range of [19, 22.5]~K in the {\it Herschel} temperature map, and is observed with signatures of star formation (including radio continuum emission). 
The JCMT $^{13}$CO(J= 3--2) line data confirm the presence of the HFS and its hub is traced with supersonic and non-thermal motions having higher Mach number and lower thermal to non-thermal pressure ratio. 
In the $^{13}$CO position-velocity diagrams, velocity gradients along the filaments toward the HFS appear to be 
observed, suggesting the gas flow in the RAFGL 5085 HFS and the applicability of the clump-fed scenario.
\end{abstract}

%\keywords{Galaxies: active---galaxies: individual (NGC 4941)---X-rays:galaxies}

\keywords{dust, extinction---HII regions---ISM: clouds---ISM: individual objects (RAFGL 5085) -- 
stars:formation}
}]

%%close the twocolumn escape here

%%include \doinum{number}for the DOI number in the header
%%include \volnum{number} for the volume number in the header
%%include \year{yyyy} for  year of publication in the header
%%include \pgrange{num--num} page range of article in the header
%%include \artcitid{num} for the article citation id
%%include \lp to print last page of the article
%%include \setcounter{page}{pagenum} for the exact starting page of the article

\doinum{12.3456/s78910-011-012-3}
\artcitid{\#\#\#\#}
\volnum{000}
\year{0000}
\pgrange{1--}
\setcounter{page}{1}
\lp{1}

\section{Introduction}
\label{sec:intro}
The role of radiative and mechanical feedback of massive OB stars ($\gtrsim$ 8 M$_{\odot}$) on the physical environment of their host galaxies has been well known in the literature \citep[e.g.,][]{zinnecker07,tan14,motte_2018}. However, despite a great deal of progress in recent years, our understanding on the origin of such massive stars is still incomplete \citep[e.g.,][]{rosen20}. 
To explain the formation of massive stars, two popular theories, which are the core-fed scenario and the clump-fed scenario, are available in the literature. The core-fed scenario (or monolithic collapse model) supports the existence of massive prestellar cores, where massive stars can form \citep{mckee_2003}. The clump-fed scenario favours that massive stars are assembled by inflow material from 
large scales of 1--10 pc clouds outside the cores \citep{bonnell_2001,bonnell_2004,bonnell_2006,vazquez_2009,vazquez_2017,vazquez_2019,padoan_2020}. In order to observationally study the formation of massive stars, one needs to explore the embedded morphology and the gas motion around a newly formed massive star that may hold clues  to its origin. 

The target site of this paper is a massive star-forming region, RAFGL 5085/IRAS 02461+6147/G136.3833+02.2666, which is located at a distance of 3.3 kpc \citep{lumsden13}. 
RAFGL 5085 has been proposed to host a massive young stellar object (MYSO) with a bolometric luminosity (L$_{bol}$) of 6580 L$_{\odot}$ \citep{lumsden13}. RAFGL 5085 is associated with very weak centimeter continuum emission \citep[i.e., total flux = 1.7 mJy;][]{NVSS} and a near-infrared (NIR) cluster \citep[e.g.,][]{carpenter00,bica03}. The NIR cluster was identified using the K$'$ stellar surface density map \citep[see Figure 3 related to IRAS 02461+6147 in][]{carpenter00}. This cluster was embedded in a molecular $^{12}$CO(1--0)/$^{13}$CO(1--0) condensation/clump \citep[at V$_{lsr}$ $\sim$$-$44 km s$^{-1}$;][]{carpenter00} traced using the FCRAO Outer Galaxy $^{12}$CO(1--0) Survey \citep{heyer98}. 
In the direction of the $^{12}$CO/$^{13}$CO clump, dense cores using the C$^{18}$O(J = 1--0) line data, and compact millimeter continuum sources (MCSs) using the continuum maps at 98 GHz and 110 GHz were reported \citep{saito06,saito07}. 
A water maser and the ammonia emission were also detected toward the NIR cluster \citep{ouyang19}. 
A bipolar molecular outflow has been investigated toward RAFGL 5085 \citep[e.g.,][]{wu04,li19}. 
Hence, previous observational studies have indicated ongoing star formation activities (including a MYSO) in the target site. 
However, embedded structure/morphology of the emission at shorter (i.e., NIR and mid-infrared (MIR)) and longer wavelengths (i.e., sub-millimetre (sub-mm)) in RAFGL 5085 is not yet carefully explored. 
Furthermore, no study is carried out to examine the gas motion toward the previously reported molecular clump hosting RAFGL 5085. 
This paper focuses on understanding the physical processes underlying star formation in RAFGL 5085 using a multi-wavelength approach, which is yet to be explored. In particular, we have used the {\it Herschel} sub-mm images at 250, 350, and 500 $\mu$m \citep[e.g.,][]{elia10,griffin10} to produce the dust column density map and the temperature map in RAFGL 5085, which are not yet studied. This paper also includes the analysis of the $^{12}$CO (J = 1--0) and $^{13}$CO (J = 3--2) line data. 

The paper is organized in the following. \S2 deals with the observational data sets used in this paper.  
We present the outcomes extracted from the analysis of multi-wavelength data sets in \S3. 
We discuss our findings in \S4. In \S5, we present the summary of the work.
\section{Data and analysis}
\label{sec:obser}
Observational data sets at different wavelengths were collected for an area of $\sim$0.22 degree $\times$ 0.22 degree 
(central coordinates: {\it l} = 136.383 degree; {\it b} = 2.255 degree) around RAFGL 5085 (see Table~\ref{tab1}).
The photometric magnitudes of point-like sources at H and K$_{s}$ bands were also obtained from the 2MASS Point Source Catalog. 
Note that the Hi-GAL observations at 70 and 160 $\mu$m and {\it Spitzer}-GLIMPSE360 map at 3.6 $\mu$m do not cover the target site. 
Hence, these data sets are not available for RAFGL 5085. This work also uses the Gaia early data release 3 \citep[EDR3;][]{gaia21,fabricius21} based photogeometric distances (``rpgeo'') of point-like sources from \citet{bailer21}.

 
We also used the bolocam source catalog at 1.1 mm \citep[v2.1;][]{ginsburg13} toward RAFGL 5085.From the JCMT Science Archive/Canadian Astronomy Data Centre (CADC), we downloaded the processed JCMT $^{13}$CO (3--2) spectral data cube (rest frequency = 330.587960 GHz) and C$^{18}$O (3--2) line data cube (rest frequency = 329.3305453 GHz) of the object ``G136.3833+02.2666/RAFGL5085" (proposal id: M08BU18), which are calibrated in antenna temperature.  
The observations (integration time = 59.358 s (for $^{13}$CO) and 59.290 s (for C$^{18}$O)) were taken using the Heterodyne Array Receiver Programme/Auto-Correlation Spectral Imaging System \citep[HARP/ACSIS;][]{buckle09} spectral imaging system. The pixel scale and the resolution of the JCMT line cube are $\sim$7\rlap.{$''$}3 and $\sim$14$''$, respectively.   
\begin{table*}
%  \tiny
\scriptsize
%\small
\setlength{\tabcolsep}{0.05in}
\centering
\caption{List of archival data sets analyzed in this paper.}
\label{tab1}
\begin{tabular}{lcccr}
\hline 
  Survey/data source  &  Wavelength/Frequency/line(s)       &  Resolution ($''$)        &  Reference \\   
\hline
NRAO VLA Sky Survey (NVSS)       & 1.4 GHz  & $\sim$45 & \citet{NVSS}\\
Milky Way Imaging Scroll Painting (MWISP)      & $^{12}$CO(J = 1--0)  & $\sim$50 & \citet{su19}\\
James Clerk Maxwell Telescope (JCMT)     & $^{13}$CO(J = 3--2), C$^{18}$O (3--2) & $\sim$14 & \citet{buckle09}\\
Bolocam Galactic Plane Survey (BGPS)       & 1.1 mm  & $\sim$33 & \citet{aguirre11}\\
{\it Herschel} Infrared Galactic Plane Survey (Hi-GAL)                              &250, 350, 500 $\mu$m     & $\sim$ 18, 25, 37         &\citet{Molinari10a}\\
Wide Field Infrared Survey Explorer (WISE) & 12, 22 $\mu$m & $\sim$6.5, $\sim$12 & \citet{wright10}\\
{\it Spitzer} Galactic Legacy Infrared Mid-Plane Survey Extraordinaire 360 (GLIMPSE360) & 4.5 $\mu$m & $\sim$2 & \citet{whitney11}\\
Two Micron All Sky Survey (2MASS)       & 1.65, 2.2 $\mu$m  & $\sim$2.5 & \citet{skrutskie06}\\
\hline          
\end{tabular}			
\end{table*}			
\section{Result}
\label{sec:result}
\subsection{Physical environments around RAFGL 5085}
\label{subsec1}
\begin{figure*}
\centering
\includegraphics[width=11.5cm]{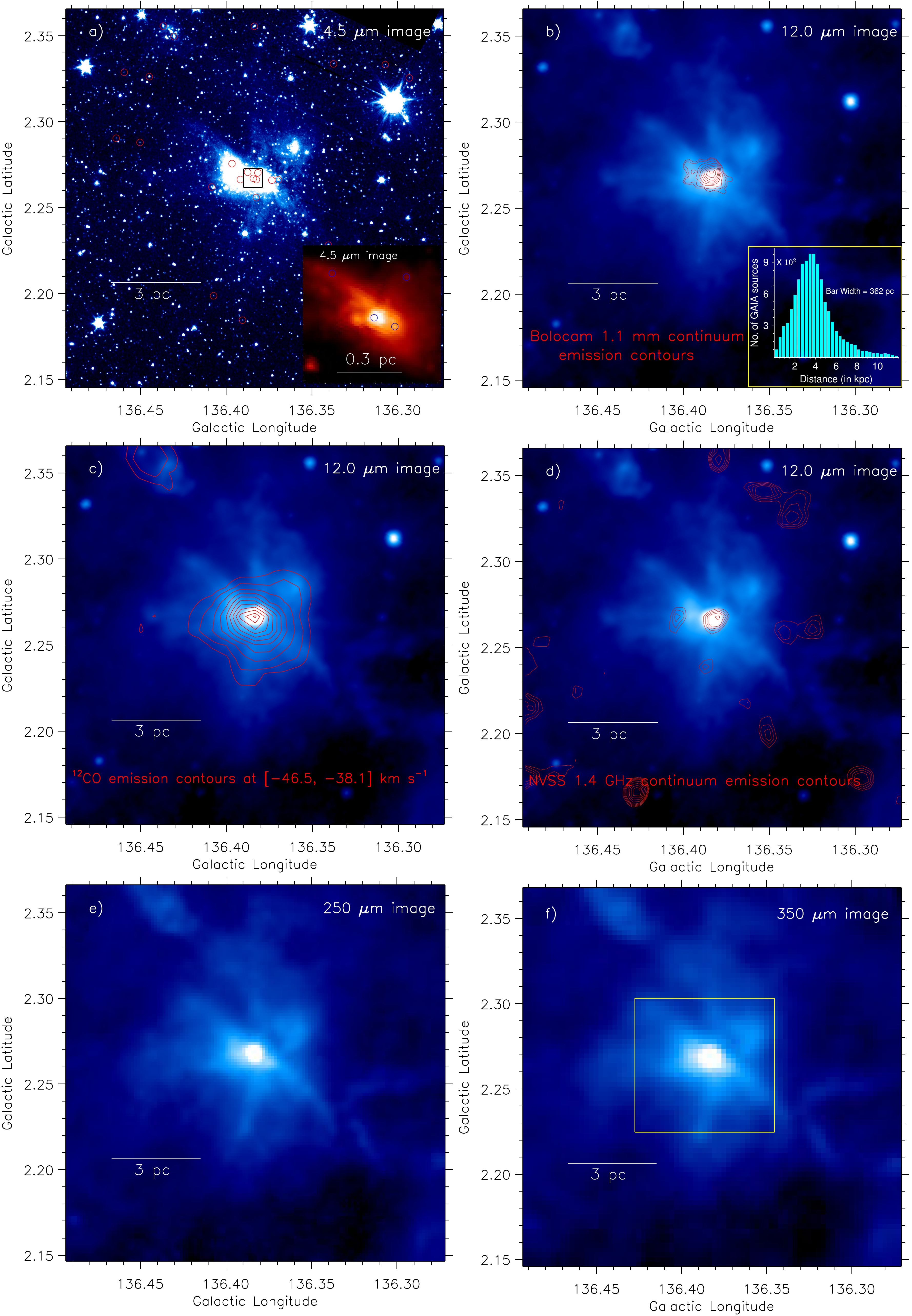}
\caption{Multi-wavelength view of RAFGL 5085. The images are shown at different wavelengths, which are highlighted in the panels. a) Overlay of infrared-excess source candidates with H$-$K$_{s}$ $>$ 0.65 (see red circles) on the {\it Spitzer} 4.5 $\mu$m image. Using the {\it Spitzer} 4.5 $\mu$m image, a zoomed-in view of the central region is shown using the inset on the bottom right (see a solid box in Figure~\ref{fig1}a). Infrared-excess source candidates are also indicated by circles. 
b) Overlay of the bolocam 1.1 mm continuum emission contours on the WISE 12 $\mu$m image, and the levels of the contours are 0.697 Jy beam$^{-1}$ $\times$ (0.25,0.3,0.4,0.5,0.6,0.7,0.8,0.9,0.98). The inset on the bottom right shows the distance distribution of Gaia point-like sources toward our selected target area. c) Overlay of the MWISP $^{12}$CO (1--0) emission contours at [$-$46.5, $-$38.1] km s$^{-1}$ on the WISE 12 $\mu$m image, and the levels of the contours are 63.1 K km s$^{-1}$ $\times$ (0.07,0.12,0.2,0.3,0.4,0.5,0.6,0.7,0.8,0.9,0.98).
d) Overlay of the NVSS 1.4 GHz radio continuum emission contours on the WISE 12 $\mu$m image, and the levels of the contours are 2.467 mJy beam$^{-1}$ $\times$ (0.35,0.4,0.45,0.5,0.6,0.7,0.8,0.9,0.98).
e) The panel shows the {\it Herschel} 250 $\mu$m image. f) The panel displays the {\it Herschel} 350 $\mu$m image.
A solid box highlights the area shown in Figures~\ref{fig2}a and~\ref{fig2}b.
In all panels, a scale bar (at a distance of 3.3 kpc) is presented.}
\label{fig1}
\end{figure*}

In the direction of the selected area, the {\it Spitzer} 4.5 $\mu$m image displays the bright extended emission or diffuse nebulosity (see Figure~\ref{fig1}a).
At least three point-like sources and an elongated feature are found toward the central region 
of the bright diffuse nebulosity, which hosts the location of RAFGL 5085 (see an inset in Figure~\ref{fig1}a).
In general, star formation activities can be inferred from the presence of infrared-excess sources and outflow activities. Such infrared-excess sources are generally identified by their varying emission at different infrared bands. Due to the absence of the {\it Spitzer} 3.6 $\mu$m image, the color conditions based on 3.6 and 4.5 $\mu$m bands cannot be used in RAFGL 5085 \citep[e.g.,][]{gutermuth09}.  Hence, we have used the 2MASS NIR data to infer the infrared-excess source candidates in our selected target area. 
In Figure~\ref{fig1}a, we have overlaid the positions of the 2MASS point-like sources having H$-$K$_{s}$ $>$ 0.65 (see open circles) on the {\it Spitzer} 4.5 $\mu$m image. This color condition is chosen based on the analysis of sources associated with a nearby control field.
Hence, the 2MASS sources with H$-$K$_{s}$ $>$ 0.65 may be considered as infrared-excess source candidates. 

Figure~\ref{fig1}b displays the WISE 12 $\mu$m image, revealing at least five filamentary features (length $>$ 1 pc) directed toward the central region hosting the location of RAFGL 5085. 
In the WISE image, a bright source is also seen toward the central region, which does not appear as a point-like source. 
The dust continuum emission at 1.1 mm is distributed toward the WISE bright source (Figure~\ref{fig1}b). We have also examined the distance distribution of Gaia point-like sources in the direction of our selected target area. \
We find a peak of distance distribution around 3.5$\pm$0.36 kpc (see an inset in Figure~\ref{fig1}b), 
which is in agreement with our adopted distance to the target source.

We find a clump in the bolocam 1.1 mm map (see Figure~\ref{fig1}b), and its total flux at 1.1 mm has been reported to be 1.824 Jy 
\citep{ginsburg13}. We have also computed the total mass of the clump using the 
following formula \citep{hildebrand83,dewangan16}:
\begin{equation}
M_{c} \, = \, \frac{D^2 \, S_\nu \, R_t}{B_\nu(T_d) \, \kappa_\nu}
\end{equation} 
\noindent where $S_\nu$ is the total flux at 1.1 mm (Jy), 
$D$ is the distance (kpc), $R_t$ is the gas-to-dust mass ratio (assumed to be 100), 
$B_\nu$ is the Planck function for a dust temperature $T_d$, 
and $\kappa_\nu$ is the dust absorption coefficient.  
The analysis uses $S_\nu$ = 1.824 Jy \citep{ginsburg13}, $\kappa_\nu$ = 1.14\,cm$^2$\,g$^{-1}$ \citep{enoch08,bally10},  
$D$ = 3.3 kpc, and $T_d$ = 22 K (see Figure~\ref{fig3x}b). 
Using these values, we have computed the mass of the bolocam clump to be $\sim$225 M$_{\odot}$. 
This $M_{c}$ value is estimated to be $\sim$255 M$_{\odot}$ at $D$ = 3.5 kpc. We assume an uncertainty on the estimate of the clump mass to be typically $\sim$20\% and at most $\sim$50\%, 
which could be contributed from the error on the adopted dust temperature, opacity, measured flux, and 
the distance of the source.

In Figure~\ref{fig1}c, using the MWISP $^{12}$CO(J = 1--0) line data, a molecular condensation is traced toward the bolocam clump, and is depicted in a velocity range of [$-$46.5, $-$38.1] km s$^{-1}$. The NVSS 1.4 GHz continuum emission is also observed toward the WISE bright source (Figure~\ref{fig1}d). 
The filamentary features become more prominent in the {\it Herschel} images at 250 and 350 $\mu$m (Figures~\ref{fig1}e and~\ref{fig1}f). 
Overall, multi-wavelength images suggest the existence of a hub-filament system (HFS; i.e., a convergence of filaments toward the compact and dense hub). Interestingly, the ionized emission and infrared-excess source candidates are evident toward the central hub of the HFS (or RAFGL 5085 HFS). 

Note that the MWISP $^{12}$CO(J = 1--0) line data do not resolve the HFS as seen in the WISE and {\it Herschel} images due to a coarse beam size. 
Hence, we examined the high resolution JCMT $^{13}$CO(J= 3--2) and C$^{18}$O(J= 3--2) line data (resolution $\sim$14$''$). 
However, the JCMT line observations do not cover the entire area as presented in Figure~\ref{fig1}a, and are available for a region highlighted by a solid box in Figure~\ref{fig1}f.
In Figure~\ref{fig2}a, we display the JCMT $^{13}$CO(J= 3--2) integrated intensity (moment-0) map (at [$-$45.35, $-$39.48] km s$^{-1}$), which reveals a central molecular condensation surrounded by parsec-scale molecular filaments. 
Figure~\ref{fig2}b shows the {\it Spitzer} 4.5 $\mu$m image overlaid with the JCMT $^{13}$CO emission contours. 
The bright diffuse nebulosity traced in the {\it Spitzer} 4.5 $\mu$m image 
is seen toward the central molecular condensation, which is surrounded by molecular filaments. 
Five arrows ``t1--t5" highlight the molecular filaments. The JCMT C$^{18}$O(J= 3--2) emission contours are also overlaid on the {\it Spitzer} 4.5 $\mu$m image, and are mainly distributed toward the central hub of the HFS. 
We do not detect any C$^{18}$O emission toward the molecular filaments as seen in the $^{13}$CO map. 
Collectively, the JCMT $^{13}$CO(J= 3--2) moment-0 map also supports the existence of the HFS.

Apart from the integrated intensity map/moment-0 map, we have also examined the intensity-weighted mean velocity/moment-1 map and the intensity-weighted velocity dispersion/intensity-weighted line width/moment-2 map of $^{13}$CO(J = 3--2).
Concerning these moment maps, one needs to collapse the data cube along the spectral axis by taking a moment of the data. The complete mathematical expressions of these moments are given in \citet{teague19}.
In Figures~\ref{fig2}c and~\ref{fig2}d, the line-of-sight velocity map (moment-1 map) of $^{13}$CO(J = 3--2) and 
the intensity-weighted Full Width Half Maximum (FWHM) line width map (moment-2) of $^{13}$CO(J = 3--2) are presented. 
Some velocity variations may be inferred toward the HFS. 
Higher line widths are evident toward the central hub compared to the filaments (see Section~\ref{subsec3}). 
All these moment maps are produced along the spectral axis from the Python package SpectralCube\footnote[1]{https://spectral-cube.readthedocs.io/en/latest/moments.html}.  
%
%Concerning moment maps, one needs to collapse the data cube along the spectral axis by taking a moment of the data
%In general, the softwares are often used to collapse the data cube along the spectral axis by taking a moment of the data  
%The complete mathematical expressions of the moment maps are given in \citet{teague19}.
%the integrated intensity; M1, the intensity-weighted average velocity, used a proxy of the line-of-sight velocity; and M2, the intensity-weighted velocity dispersion,
%A common practice in the analysis of line emission is to collapse the data cube along the spectral axis by taking a moment of the data
%
%This method does not depend on profile fitting, it provides a robust estimate of velocity.
%The velocity map and the line width map were made by fitting a 1D Gaussian to each pixel's spectrum along the velocity axis. 
%These exercises are performed using the Python package.} 
%
\subsection{Identification of filament skeletons and their configuration}
\label{hfs:iden}
In order to reveal the structures of filaments or filamentary skeletons, we employed the ``getsf" \citep{getsf_2022} algorithm on the {\it Herschel} 250 $\mu$m image. This particular image has better spatial resolution compared to other {\it Herschel} images 
at 350 and 500 $\mu$m. The algorithm distinguishes the structural components from their backgrounds in a given astronomical image.
It requires the input of maximum width of filament (i.e., FWHM in arcsec), which the user wants 
to extract \citep[see][for the detailed steps]{bhadari22}.
However, the output files include the scale dependent (i.e., from the image resolution scale to the maximum width of filaments) skeletons within the range of the user's scale of interest. The filament skeletons identified on the scale of 51$''$ are presented in Figure~\ref{fig3}a. 
In order to run the ``getsf", we set the maximum filament width size of 60$''$.
Figure~\ref{fig3}b is the same as Figure~\ref{fig3}a, but it also shows the $^{13}$CO(J= 3--2) emission contours (see Figure~\ref{fig2}b). In Figures~\ref{fig3}a and~\ref{fig3}b, the HFS and its association with molecular gas are evident.
\subsection{{\it Herschel} column density and temperature maps}
\label{subsec2}
Figures~\ref{fig3x}a and~\ref{fig3x}b show the {\it Herschel} column density and temperature maps (resolution $\sim$37$''$), respectively.
Following the methods described in \citet{mallick15}, these {\it Herschel} maps are produced using the Hi-GAL continuum images at 250, 350, and 500 $\mu$m.
Note that the {\it Herschel} 160 $\mu$m image is not used in the analysis due to its unavailability.
The {\it Herschel} column density and temperature maps are generated from a pixel-by-pixel spectral energy distribution (SED) fit with a modified blackbody to the sub-mm emission at 250--500 $\mu$m. A background flux level was estimated for each sub-mm wavelength, and is determined to be 0.18, 0.13, and 0.06 Jy/pixel for the 250, 350, and 500 $\mu$m images (size of the selected featureless dark region $\sim$9$'$ $\times$ 9$'$; centered at: {\it l} = 135.412 deg;
{\it b} = 2.092 deg), respectively. 
In this analysis, we considered a mean molecular weight per hydrogen molecule (${\mu }_{{\rm{H}}2}$) of 2.8 \citep{kauffmann08}, an absorption coefficient (${\kappa }_{\nu }$) of 0.1 ${(\nu /1000\mathrm{GHz})}^{\beta }$ cm$^2$\,g$^{-1}$ with a dust spectral index ($\beta$) of 2 \citep[see][]{hildebrand83}, and a gas-to-dust ratio of 100.
The HFS is not very prominent in the {\it Herschel} column density map, but the central hub is associated with warm dust emission ($T_d$ $\sim$19--22.5~K). Uncertainties on the estimate of the column densities and the dust temperatures could be $\sim$10--20\% \citep[e.g.,][]{launhardt13}.

Figure~\ref{fig3x}c presents filaments identified by ``getsf" on the {\it Herschel} 250 $\mu$m image (see Figure~\ref{fig3}a), which are utilized to mask the column density map as shown in Figure~\ref{fig3x}a.
Figure~\ref{fig3x}c exhibits filaments with high aspect ratios (length/diameter)
and lower column densities ($\sim$0.1--2.4 $\times$10$^{21}$ cm$^{-2}$), while
the central hub is found with a low aspect ratio and high column density ($\sim$3.5--7.0 $\times$10$^{21}$ cm$^{-2}$). Figure~\ref{fig3x}d is the same as Figure~\ref{fig3x}c, but it also displays the $^{13}$CO(J= 3--2) emission contours (see Figure~\ref{fig2}b).
We find the spatial connection of molecular gas with the structures seen in the column density map (see Figure~\ref{fig3x}d). 
\subsection{JCMT position-velocity diagrams and non-thermal velocity dispersion map}
\label{subsec3}
\begin{figure*}
\centering
\includegraphics[width=\textwidth]{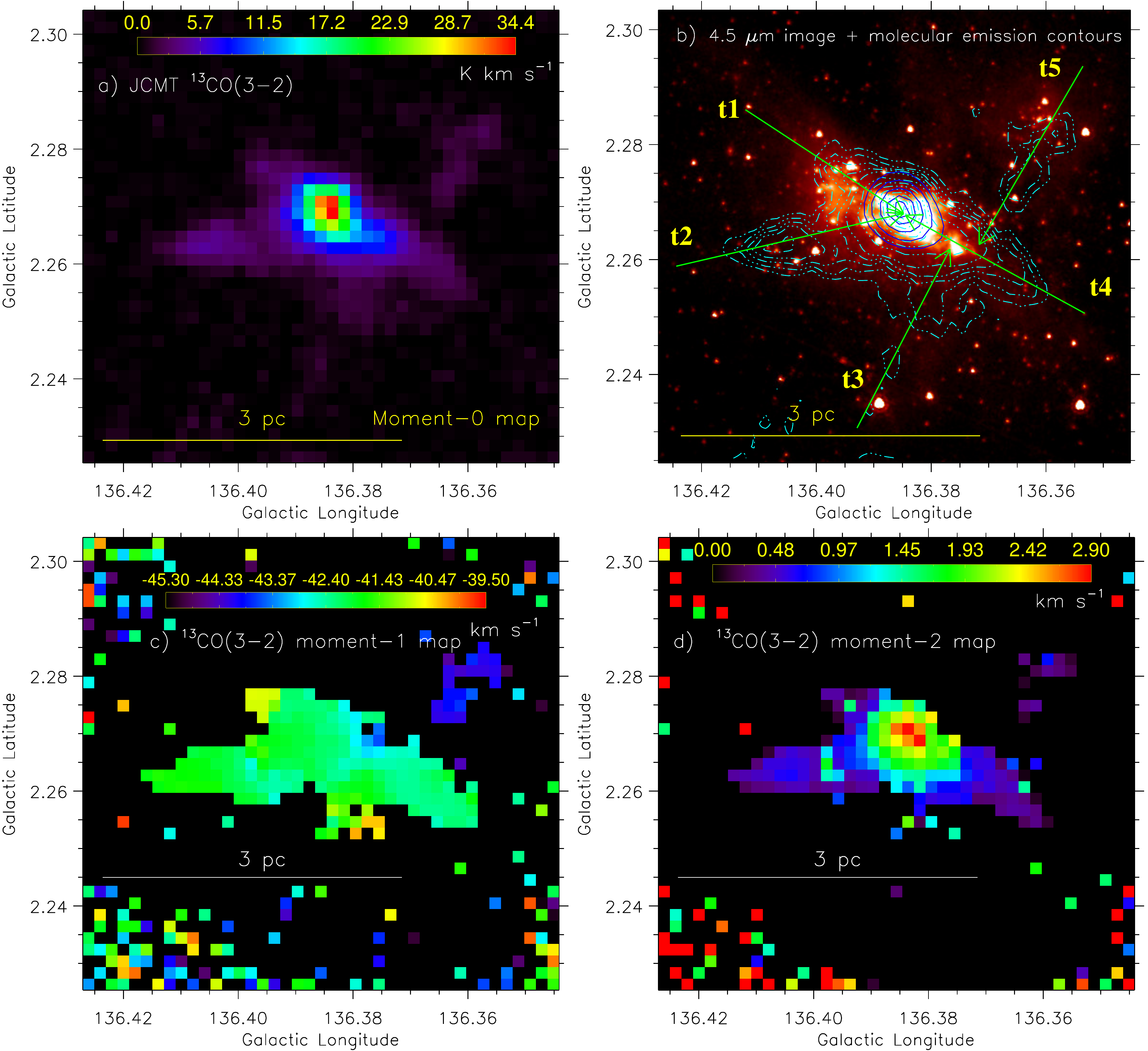}
\caption{a) The panel shows the JCMT $^{13}$CO(J= 3--2) integrated intensity (moment-0) map (at [$-$45.35, $-$39.48] km s$^{-1}$) of an area hosting the target site (see a box in Figure~\ref{fig1}f).
b) Overlay of the $^{13}$CO(J= 3--2) emission contours on the {\it Spitzer} 4.5 $\mu$m image.
The $^{13}$CO(J= 3--2) contours (in cyan) are 34.13 K km s$^{-1}$ $\times$ (0.03,0.05,0.07,0.09,0.12,0.2,0.3,0.4,0.5,0.6,0.7,0.8,0.9,0.98). Solid blue contours of the JCMT C$^{18}$O(J= 3--2) integrated intensity map (at [$-$45.35, $-$39.48] km s$^{-1}$) are also presented with the levels of 1, 1.5, 2.1, 2.6, and 2.9 K km s$^{-1}$. Five arrows ``t1--t5" are marked in the panel, where the position-velocity diagrams are generated
(see Figures~\ref{fig4cc}a--\ref{fig4cc}e). c) The panel shows the line-of-sight velocity map (moment-1 map) of the JCMT $^{13}$CO(J= 3--2) emission. 
d) JCMT $^{13}$CO(J= 3--2) intensity-weighted FWHM line width map (moment-2 map). A scale bar corresponding to 3 pc (at a distance of 3.3 kpc) is shown in each panel.}
\label{fig2}
\end{figure*}

\begin{figure}
\centering
\includegraphics[width=8.5cm]{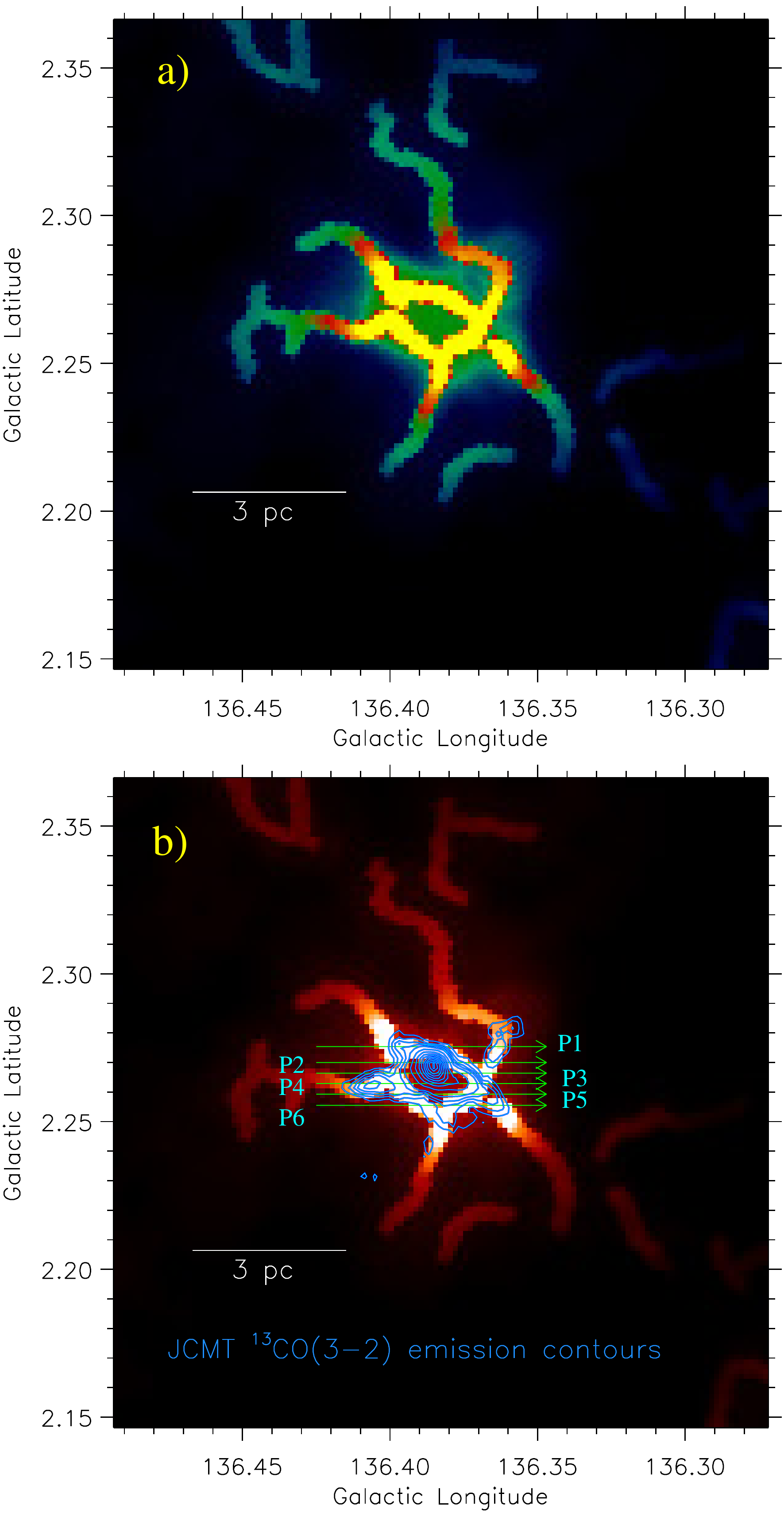}
\caption{a) The panel shows the emission skeletons (on the scale of 51$''$) identified from the {\it Herschel} 250{\mic} image (resolution $\sim$18$''$) using the algorithm ``getsf".
b) Same as Figure~\ref{fig3}a, but it is overlaid with the $^{13}$CO(J= 3--2) contours (see Figure~\ref{fig2}b).
Six arrows ``p1--p6" are indicated in the panel, where the position-velocity diagrams are generated
(see Figures~\ref{fig4}a--\ref{fig4}f). A scale bar corresponding to 3 pc (at a distance of 3.3 kpc) is shown in each panel.}
\label{fig3}
\end{figure}

As mentioned earlier, the JCMT molecular line data do not cover the entire HFS as traced in the {\it Herschel} maps, and are available for the central area of the HFS (see Figure~\ref{fig3}b). Hence, to examine the gas motion toward the HFS, we have produced position-velocity diagrams along several paths passing through the molecular filaments and the central hub (see five arrows t1--t5 in Figure~\ref{fig2}b and six horizontal arrows p1--p6 in Figure~\ref{fig3}b). 
The arrows t1--t5 are selected toward the molecular filaments, and in such paths, we take the outermost edge of each filament as the reference point and go towards the central hub. On the other hand, the horizontal arrows p1--p6 pass through the central hub and some molecular filaments. 
Hence, the paths t1--t5 and the paths p1--p6 may pass through the common areas of the HFS.  

Position-velocity diagrams along arrows t1--t5 are presented in Figures~\ref{fig4cc}a--\ref{fig4cc}e. These diagrams hint at the presence of a noticeable velocity gradient along the molecular filaments (see Figures~\ref{fig4cc}b,~\ref{fig4cc}d, and~\ref{fig4cc}e). 
In the direction of the filament ``t2", a velocity gradient is determined to be about 1.6 km s$^{-1}$ pc$^{-1}$. 
In Figures~\ref{fig4}a--\ref{fig4}f, we also display the position-velocity diagrams along arrows p1--p6. 
The arrows p2--p4 seem to pass through the central hub of the HFS (see also the arrow ``t2" in Figure~\ref{fig2}b), and the position-velocity maps along these paths suggest the outflow activity associated with
the central hub (see Figures~\ref{fig4}b--\ref{fig4}d), where infrared-excess source candidates are observed.
Two arrows p4 and p5 appear to pass through the molecular filaments of the HFS, and one can see a noticeable velocity gradient along the molecular filaments toward the HFS in the position-velocity maps (see Figures~\ref{fig4}d,~\ref{fig4}e, and also Figure~\ref{fig4cc}b). 
Considering all these maps, there is a hint of the presence of the velocity gradient toward the HFS. To further carry out such study, high resolution molecular line observations for a wide area around RAFGL 5085 will be required. 

\begin{figure*}
\centering
\includegraphics[width=\textwidth]{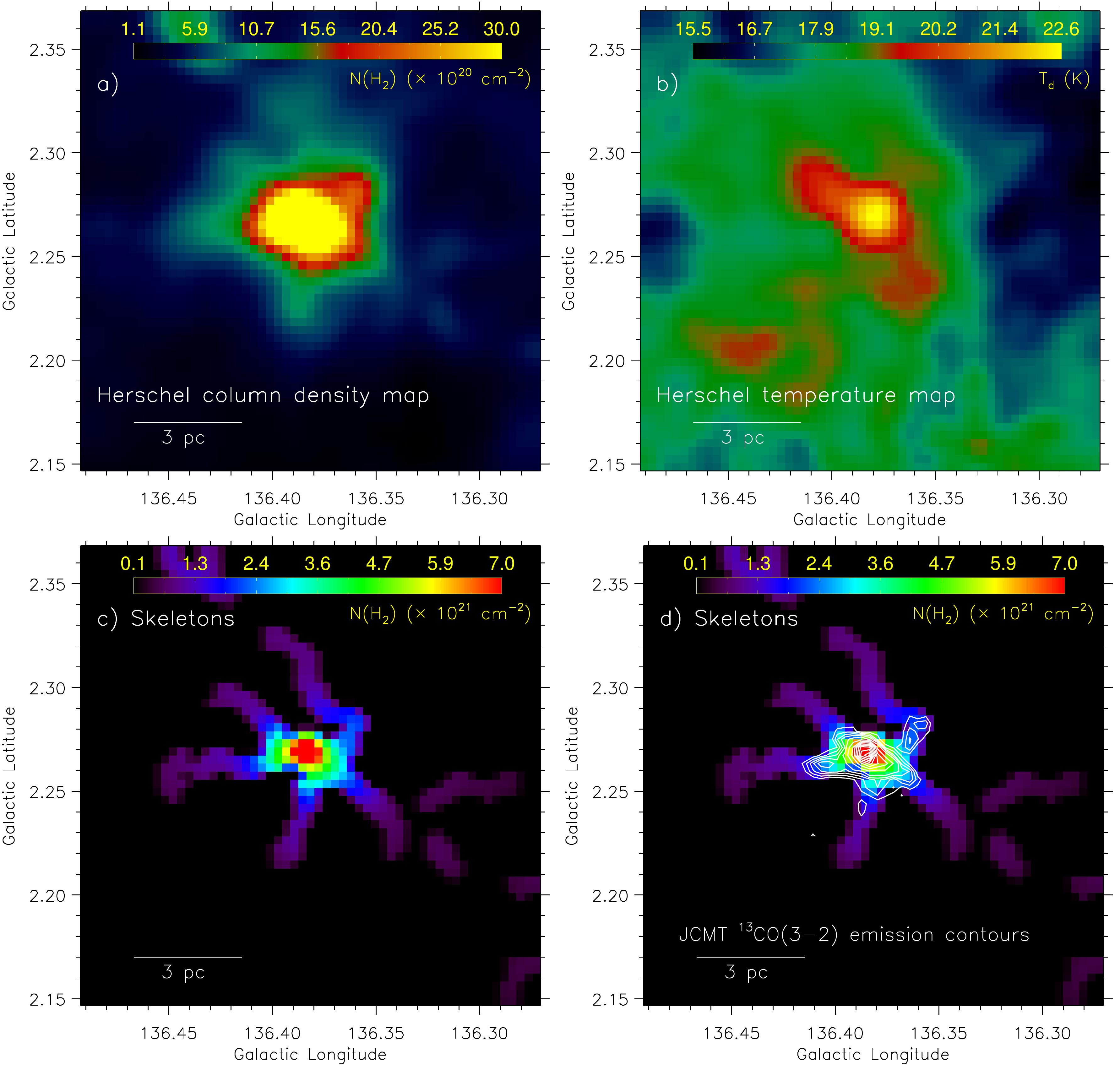}
\caption{a) The panel displays the {\it Herschel} column density map (resolution $\sim$37$''$) of an area presented in Figure~\ref{fig1}a.
b) The panel shows the {\it Herschel} temperature map (resolution $\sim$37$''$).
c) The panel displays filaments identified on the {\it Herschel} 250 $\mu$m by the algorithm ``getsf" \citep{getsf_2022},
which are utilized to mask the column density map as shown in Figure~\ref{fig3x}a.
d) Same as Figure~\ref{fig3x}c, but it is overlaid with the $^{13}$CO(J= 3--2) contours (see Figure~\ref{fig2}b).
A scale bar corresponding to 3 pc (at a distance of 3.3 kpc) is shown in each panel.}
\label{fig3x}
\end{figure*}

\begin{figure*}
\centering
\includegraphics[width=\textwidth]{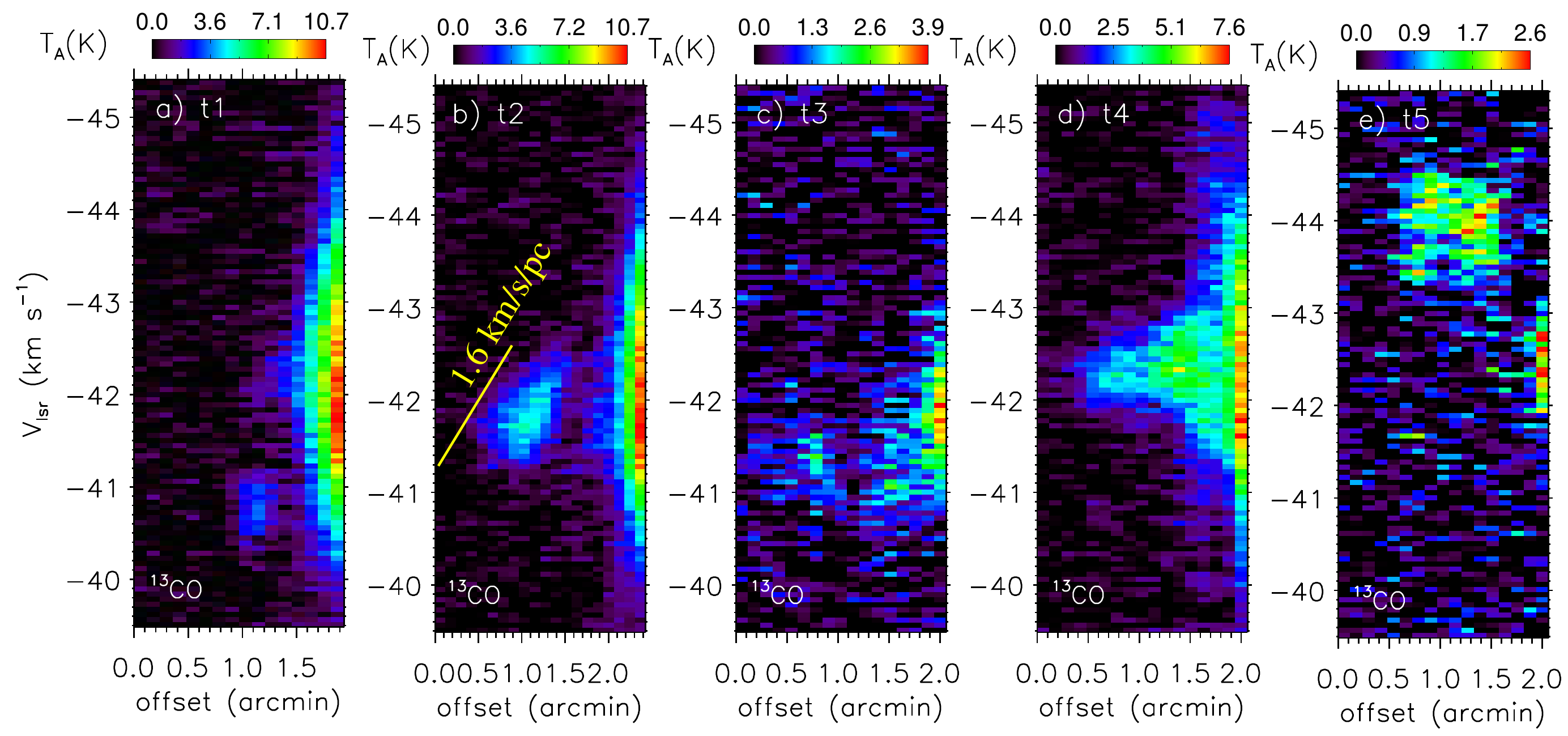}
\caption{Position-velocity diagrams of the $^{13}$CO(J= 3--2) emission along arrows a) ``t1"; b) ``t2"; c) ``t3"; d) ``t4"; e) ``t5" (see arrows in Figure~\ref{fig2}b). In panel ``b", a reference bar at 1.6 km s$^{-1}$ pc$^{-1}$ is marked to show a velocity gradient.}
\label{fig4cc}
\end{figure*}

\begin{figure*}
\centering
\includegraphics[width=\textwidth]{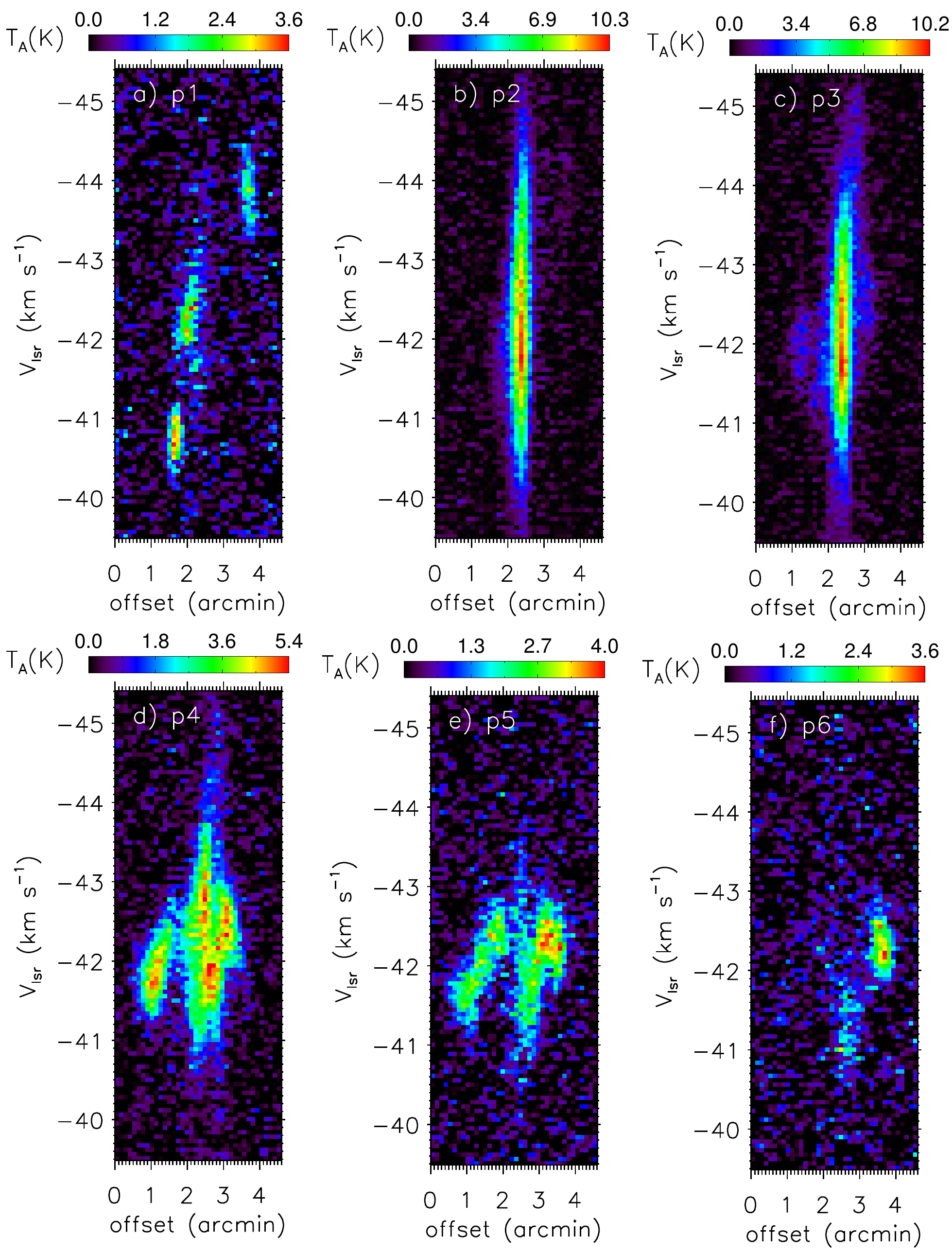}
\caption{Position-velocity diagrams of the $^{13}$CO(J= 3--2) emission along arrows a) ``p1"; b) ``p2"; c) ``p3"; d) ``p4"; e) ``p5"; f) ``p6")  (see arrows in Figure~\ref{fig3}b).}
\label{fig4}
\end{figure*}

We have also utilized the JCMT $^{13}$CO(J = 3--2) line data to infer the
non-thermal velocity dispersion, sound speed, Mach number, ratio of thermal to non-thermal
pressure toward the molecular features traced in the JCMT moment-0 map.
An expression of the non-thermal velocity dispersion is defined below:
\begin{equation}
\sigma_{\rm NT} = \sqrt{\frac{\Delta V^2}{8\ln 2}-\frac{k T_{kin}}{29 m_H}} = \sqrt{\frac{\Delta V^2}{8\ln 2}-\sigma_{\rm T}^{2}} ,
\label{sigmanonthermal}
\end{equation}
where $\Delta V$ is the measured FWHM line width, $\sigma_{\rm T}$ (= $(k T_{kin}/29 m_H)^{1/2}$) is the thermal broadening for $^{13}$CO at gas kinetic temperature (T$_{kin}$), and $m_H$ is the mass of hydrogen atom.
Mach number can also be determined by taking the ratio of non-thermal velocity dispersion ($\sigma_{\rm NT}$) to sound speed ($a_{s}$).
An expression of the sound speed is $a_{s}$ = $(k T_{kin}/\mu m_{H})^{1/2}$, where $\mu$ is the mean molecular weight ($\mu$=2.37; approximately 70\% H and 28\% He by mass). Such estimates also allow us to compute the ratio of thermal to non-thermal (or turbulent) pressure \citep{lada03}, and its expression is $R_{p} = {a_s^2}/{\sigma^2_{NT}}$. Note that we do not have the knowledge of T$_{kin}$.
Hence, dust temperature is used in the calculation, and is taken from the {\it Herschel} temperature map (see Figure~\ref{fig3x}b).

\begin{figure*}
\centering
\includegraphics[width=\textwidth]{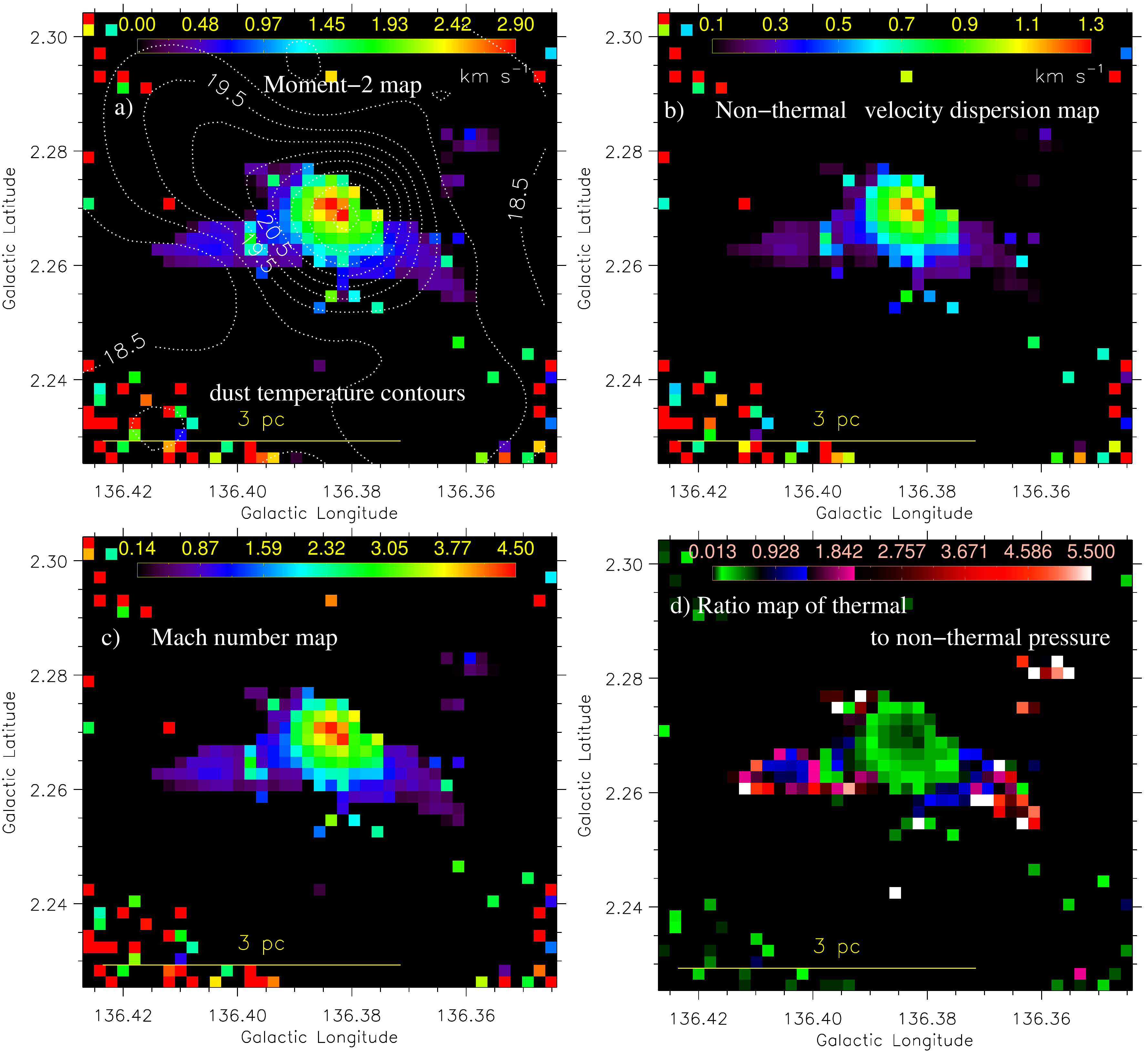}
\caption{a) JCMT $^{13}$CO(J= 3--2) moment-2 map (see also Figure~\ref{fig2}d). The contours of dust temperature at [18.5, 19.0, 19.5, 20.0, 20.5, 21.0, 21.5, 22.0, 22.5]~K are also overlaid on the moment-2 map (see also Figure~\ref{fig3x}b).
b) JCMT $^{13}$CO(J= 3--2) non-thermal velocity dispersion map.
c) JCMT $^{13}$CO(J= 3--2) Mach number map.
d) JCMT $^{13}$CO(J= 3--2) ratio map of thermal to non-thermal (or turbulent) pressure.}
\label{fig6x}
\end{figure*}

In Figure~\ref{fig6x}a, we display the moment-2 map of $^{13}$CO(J = 3--2) overlaid
with the contours of dust temperature at [18.5, 19.0, 19.5, 20.0, 20.5, 21.0, 21.5, 22.0, 22.5]~K. 
Figures~\ref{fig6x}b,~\ref{fig6x}c, and~\ref{fig6x}d show the JCMT $^{13}$CO(J= 3--2) $\sigma_{\rm NT}$ map, Mach number map, and R$_{p}$ map, respectively. In the direction of the central hub, the values of $\Delta V$, $\sigma_{\rm NT}$, and Mach number are larger than other molecular features. Furthermore, a smaller value 
of R$_{p}$ ($<$1) is found toward the central hub, suggesting that the non-thermal pressure is higher than the thermal pressure in the hub.  
\begin{figure*}
\centering
\includegraphics[width=\textwidth]{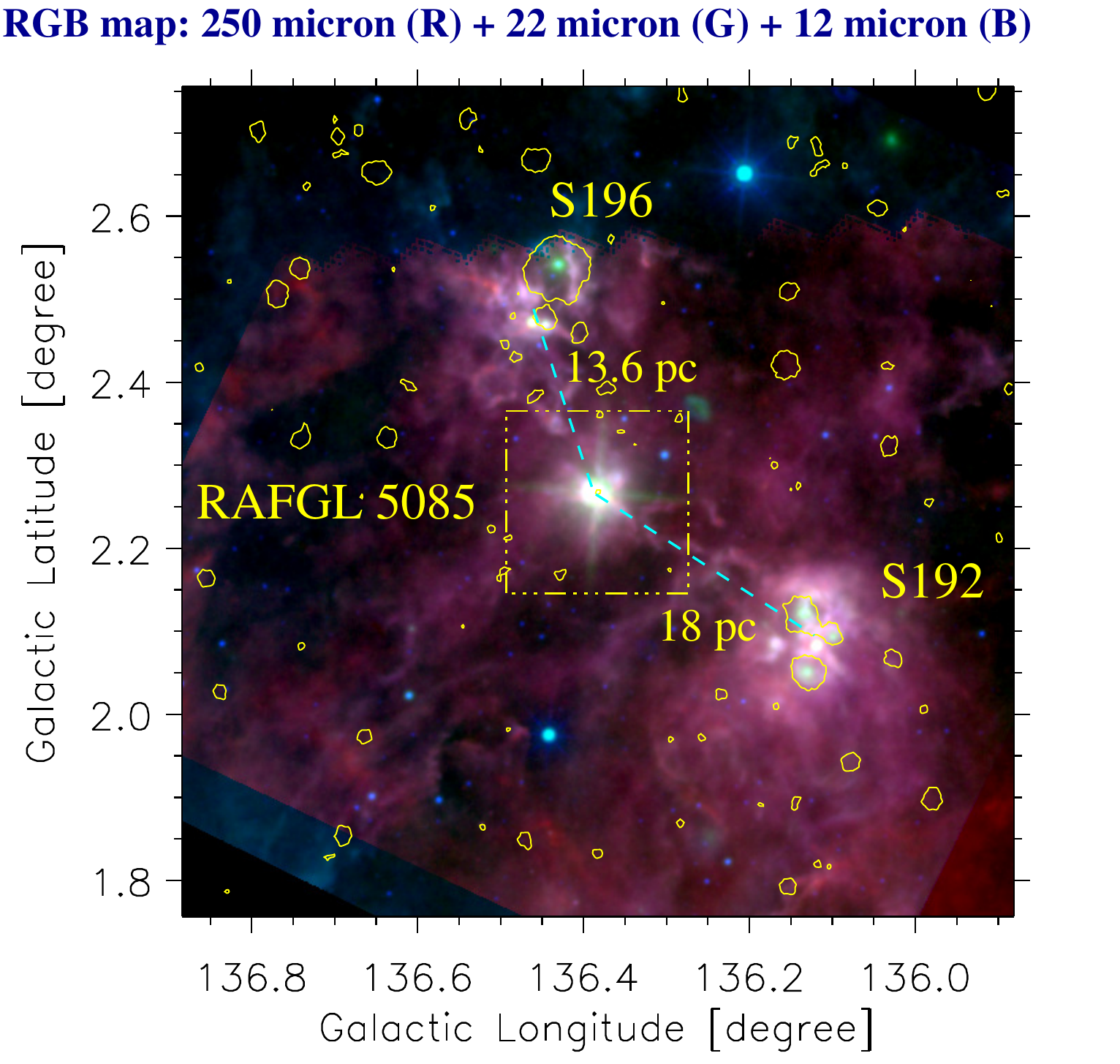}
\caption{Large-scale view of an area containing the site RAFGL 5085. The panel displays a 3-color composite map made using the {\it Herschel} 250 $\mu$m (in red), WISE 22 $\mu$m (in green), and WISE 12 $\mu$m (in blue) images. The NVSS 1.4 GHz radio continuum emission contour at 1.35 mJy beam$^{-1}$ is also overlaid on the color composite map. The target area of this paper is indicated by a dotted-dashed box.}
\label{fig7x}
\end{figure*}
\section{Discussion}
\label{sec:disc}
To examine a wider scale environment of our selected target area containing the site RAFGL 5085, we present a color-composite map (250 $\mu$m (in red), 22 $\mu$m (in green), and 12 $\mu$m (in blue) images), which is also overlaid with the NVSS 1.4 radio continuum emission. The H\,{\sc ii} regions associated with the sites S192 and S196 are seen, and their physical separations from our target site RAFGL 5085 are also indicated in Figure~\ref{fig7x}. 
The extended MIR emission is traced toward the sites S192 and S196, while the compact MIR emission is found toward RAFGL 5085. 
From Figure~\ref{fig7x}, we notice that RAFGL 5085 appears to be located between the two evolved H\,{\sc ii} regions S192 and S196. 

Previously, the NIR cluster -- associated with the outflow signature, massive star formation and water maser -- has been reported toward the molecular and dust clump hosting the location of RAFGL 5085 \citep[][]{carpenter00,wu04,saito06,saito07,lumsden13,li19,ouyang19}, where dense molecular cores have been identified \citep[e.g.,][]{saito06}. Due to the detection of weak radio continuum emission \citep[i.e., total flux = 1.7 mJy;][]{NVSS}, we do not favour the presence of any evolved H\,{\sc ii} region in RAFGL 5085. Hence, the site RAFGL 5085 seems to be associated with an early stage of massive star formation \citep[see also][]{lumsden13}. In this paper, using the {\it Herschel} sub-mm images and the WISE image at 12 $\mu$m, our findings reveal a new picture in RAFGL 5085, which is the existence of a HFS.
In this connection, the algorithm ``getsf" identified at least five filaments (having $N({{\rm{H}}}_{2})$ $\sim$0.1--2.4 $\times$10$^{21}$ cm$^{-2}$), which appear to direct to the central hub (having $N({{\rm{H}}}_{2})$ $\sim$3.5--7.0 $\times$10$^{21}$ cm$^{-2}$).
This finding is also supported by the JCMT $^{13}$CO(J= 3--2) line data.
All the signatures of star formation (including radio continuum emission) are found toward the central hub of the HFS (see Section~\ref{subsec1}), where higher values of Mach number and non-thermal pressure are depicted (see Section~\ref{subsec3}). 

Prior to the {\it Herschel} era, HFSs hosting protoclusters were studied by \citet{myers_2009}.
After the availability of the {\it Herschel} sub-mm maps, in order to understand the process
of mass accumulation in MSF, the study of HFS and its implication has received considerable attention \citep[e.g.,][]{motte_2018}.
In hub-filament configurations, it is thought that filaments are instrumental in channelling interstellar gas and dust to the central hub, where massive stars are assembled by the inflow material \citep{andre_2010,schenider_2012,motte_2018,morales19,rosen20}. 
In this context, velocity gradients (e.g., 0.5--2.5 km s$^{-1}$ pc$^{-1}$) along molecular filaments have been observed in HFSs, and are suggested as a signature of gas accretion along the filaments \citep{kirk13,nakamura14,olmi16,hacar17,baug18,morales19,chen20,dewangan20t}. Such material/gas flow is thought to feed star-forming cores and proto-clusters in the hub \citep[e.g.,][]{morales19}. The earlier reported velocity gradients in the filaments were determined to be higher than the sound speed of $\sim$0.2 km s$^{-1}$ at T = 10 K, indicating a flow of turbulent gas \citep[e.g.,][]{baug18}. The presence of a velocity gradient toward the HFS is found (see Figures~\ref{fig4cc} and~\ref{fig4}), hinting gas flow along filaments to the central hub. 
 However, we do not have enough molecular line data for the entire HFS to further probe the velocity gradients \citep[e.g.,][]{hacar11,hacar13}. Hence, this demands new high-resolution and high-sensitivity molecular line data for a wide area around the RAFGL 5085 HFS to further confirm our proposed argument.  
The detection of a hub-filament configuration supports the applicability of the clump-fed scenario as discussed in the Introduction section.  In the clump-fed scenario, inflow material from very large scales of 1--10 pc is thought to be responsible for the birth of massive stars. 
Such inflow material can be gravity driven or are produced due to supersonic turbulence \citep[e.g.,][]{motte_2018,padoan_2020,hongli22}. 
With the available data sets mentioned in Table~\ref{tab1}, the detailed study of this proposed aspect is outside the scope of this paper.

Based on gravity driven inflow, an evolutionary scheme has been proposed to explain the birth of massive stars \citep{tige_2017,motte_2018}. This scheme does not favour the existence of high-mass prestellar cores.
Rather, low-mass prestellar cores are suggested to develop first (within $\sim$10$^{4}$ years) in the starless phase of massive dense cores/clumps (MDCs, in a 0.1~pc scale). 
The protostellar phase of MDCs begins when the low-mass protostellar core forms. The local ($\sim$0.02 pc) collapse of these cores is accompanied by the global ($\sim$0.1--1 pc) collapse of MDCs and hub region. In this way, the low-mass protostellar core gains mass from these gravity driven inflows and becomes a high-mass protostar ($\sim$3$\times$10$^{4}$ years). 
In the later stage, the scheme favours the development of H\,{\sc ii} regions around massive protostars with masses
greater than 8 M$_{\odot}$. 
The H\,{\sc ii} region forms when the strong UV field from a massive protostar ionizes the protostellar envelope. Simultaneously the main accretion phase also terminates.
According to this scheme, one may expect simultaneous growth of stars, cores, and ridges from the mass of their parental cloud.

The HFSs are very common structures in the massive star-forming sites, but a perfect hub-filament configuration is not often seen in most cases. This is because of the stellar feedback from H\,{\sc ii} regions (or proto clusters) which gradually destroys the HFS as the proto cluster evolves \citep[see][]{baug15,baug18,dewangan17s,dewangan20t,dewangan22a,morales19,dewangan21a,dewangan22ex}. It seems that the observed RAFGL 5085 HFS is not yet influenced by UV radiation from the H\,{\sc ii} region. 
Hence, our target source RAFGL 5085 appears to be one of the perfect HFS candidates where the young sources (including massive protostar) are found to be formed in the hub region. 
\section{Conclusion and Summary}
\label{sec:conc}
In this paper, using a multi-wavelength approach, the physical environment of a massive star-forming site RAFGL 5085 has been studied. 
Our visual inspection of the continuum images at 12, 250, 350, and 500 $\mu$m has revealed a hub-filament configuration, 
which consists of a central region (M$_{\rm clump}$ $\sim$225 M$_{\odot}$) surrounded by at least five parsec-scale filaments.  
We also applied the getsf tool in the {\it Herschel} sub-mm images to extract filaments, and their observed configuration supports 
the presence of the HFS in the site RAFGL 5085. In the {\it Herschel} column density map, filaments are identified with higher aspect ratios (length/diameter)
and lower $N({{\rm{H}}}_{2})$ values ($\sim$0.1--2.4 $\times$10$^{21}$ cm$^{-2}$), while the central hub is found with a lower aspect ratio and higher $N({{\rm{H}}}_{2})$ values ($\sim$3.5--7.0 $\times$10$^{21}$ cm$^{-2}$). The central hub of the RAFGL 5085 HFS is associated with warmer dust emission (i.e., T$_{\rm d}$ $\sim$[19, 22.5]~K). Signposts of star formation (including radio continuum emission) are concentrated toward the central hub. Our analysis of the JCMT $^{13}$CO(J= 3--2) line data also confirms the presence of the HFS in RAFGL 5085. One can note that the JCMT molecular line data are not available for the entire HFS as found in the {\it Herschel} maps, and cover mainly the central area of the HFS. The central hub is traced with supersonic and non-thermal motions having higher Mach number and lower thermal to non-thermal pressure ratio. 
In the $^{13}$CO(J= 3--2) position-velocity diagrams, velocity gradients along the filaments toward the HFS seem to be present, suggesting the gas flow in the RAFGL 5085 HFS and the applicability of the clump-fed scenario. We also suggest that the site RAFGL 5085 hosts the early phase of massive star formation, where the presence of an evolved H\,{\sc ii} region is unlikely. 
Hence our selected target site appears to be a perfect HFS candidate, which is not yet affected by UV photons from the H\,{\sc ii} region. 
%
%\vspace{2em}
%\vspace{-2em}
%
\section*{Acknowledgements}
We are thankful to the anonymous reviewer for the useful comments and suggestions.  
The research work at Physical Research Laboratory, Ahmedabad, is funded by the Department of Space, Government of India. 
This research made use of the data from the Milky Way Imaging Scroll Painting (MWISP) project, which is a multi-line survey in 12CO/13CO/C18O along the northern galactic plane with PMO-13.7m telescope. We are grateful to all the members of the MWISP working group, particularly the staff members at PMO-13.7m telescope, for their long-term support. MWISP was sponsored by National Key R\&D Program of China with grant 2017YFA0402701 and by CAS Key Research Program of Frontier Sciences with grant QYZDJ-SSW-SLH047.
This research used the facilities of the Canadian Astronomy Data Centre operated 
by the National Research Council of Canada with the support of the Canadian Space Agency. C.E. acknowledges the financial support from grant RJF/2020/000071 as a part of the Ramanujan Fellowship awarded by the Science and Engineering Research Board (SERB), Department of Science and Technology (DST), Govt. of India.

%%Use section* for acknowledgements
%\section*{Acknowledgements}
\vspace{-1em}

%%use \balance somewhere in the left column of the last page to balance the two columns in the end page

%%References section

%%%%%%%%%%%%%%%%%%%% REFERENCES %%%%%%%%%%%%%%%%%%

\bibliographystyle{apj}
\bibliography{reference}
%{}
%\nocite{*}

%%%%%%%%%%%%%%%%%%%%%%%%%%%%%%%%%%%%%%%%%%%%%%%%%%

%\end{theunbibliography}

\end{document}